\def\be{\begin{equation}}
\def\ee{\end{equation}}
\documentstyle[aps,preprint,epsf]{revtex}
\begin{document}
\draft
\title{Phase space diffusion and low temperature aging}
\author{A. Barrat and M. M\'ezard}

\address{Laboratoire de Physique Th\'eorique de l'Ecole
Normale Sup\'erieure\footnote{Unit\'e propre du CNRS, associ\'ee \`a
 l'Ecole Normale Sup\'erieure et \`a l'Universit\'e de
Paris Sud}, 24 rue Lhomond, 75231 Paris Cedex 05, France}
\date{\today}
\maketitle
\begin{abstract}
We study the dynamical evolution of a system with a
phase space consisting of configurations with random energies.
The dynamics we use is of Glauber type. It allows for some
dynamical evolution ang aging even at very low temperatures,
through the search of configurations with lower energies.
\end{abstract}
\vskip 1cm
\begin{center}
LPTENS preprint 95/16
\end{center}


Submitted to:

{\it Journal de Physique I}

2.Statistical physics
\newpage
Lots of efforts are being devoted to the study of
the out of equilibrium dynamics of spin-glasses, in
order to understand several experimental findings, like aging,
and the joint existence of dynamics and memory in temperature cycling
experiments \cite{exp}. Various approaches have been used, including
some phenomenological studies based on
droplet or domain growth \cite{droplets},
 numerical simulations \cite{Rieger},
mean-field models
\cite{CuKu,Fr.MM,CuKu2}, diffusion on tree like structures \cite{tree}, and
 models based on the existence of traps in phase space.

In this note we shall elaborate along the lines of this last approach.
It deals directly with the structure
of the phase space, made of many metastables states, figuring the
configurations of a spin system.
Such a picture has been put forward in \cite{Bouchaud1}, and
generalized in \cite{Bouchaud3}, presenting the
phase space as a random energy landscape made of traps, with a broad
distribution of trapping times: the energies are the low
lying ones of a REM \cite{derr}, so they have an exponential
probability distribution, and the trapping times, given by an
Arrhenius law, have a
power-law distribution with infinite mean. The dimension of
the space is infinite (equivalently, all the traps are connected).
In this frame, the diffusion is anomalous \cite{Bouchaud1,Bouchaud3,Bouchaud2},
and aging is present. One virtue of this approach has been to point out a
simple ``kinematic'' ingredient which induces aging: in a system
with a broad distribution of trapping times, the probability to be at time
$t$ in a trap of lifetime $\tau$ always depends on $t$, and this induces aging.
Besides this kinematic effect, there might well exist  another,
``dynamical'', source of aging,
namely the explicit evolution with time of the distribution of trapping
times. We shall provide hereafter such an example of a dynamical
aging process.

In the trap model, the probability of hopping from one configuration
 $i$ to another
configuration $j$, $W_{i \rightarrow j}$, depends only on the energy of the
 $i$: the energies are in fact seen as energy barriers, which are
then uncorrelated. Another
case was studied in \cite{ddol}, where $W_{i \rightarrow j}$ depends
only of the arrival configuration $j$, which allows a great simplification
of the master equation. Only these two extreme cases have been studied so
far. Hereafter we shall consider
a case where  the
 transition probability, which is of Glauber type, depends
on both configurations, so the energy barriers are now correlated: this
seems a priori a more refined way to define a dynamic for the REM.
Furthermore this case allows
for the existence of a dynamical evolution even at zero temperature, through
the search of configurations with lower energies. During this
evolution it becomes more and more difficult for the system  to find a lower
configuration, which results in a slow down of the dynamics. We
shall show through an
explicit solution at zero temperature that this mechanism gives rise to
an aging effect in which the system never reaches equilibrium.
This aging  does
not take its origin in energetic barriers, but rather in some
kind of entropic barrier, namely the low probability of finding
a favourable direction in phase space. Another example of aging due
to entropic barrier has been proposed recently by Ritort \cite{ritort}.

The model we consider is defined as follows: The system can be in
any of
 $N$ configurations $i=1,..,N$. The configuration $i$ has an  energy $E_{i}$.
The
energies are independent random
variables
with distribution $P(E)$.
The probability of hopping from one configuration to another can be defined
in several ways, the only a priori constraint being the detailed
balance:
\begin{equation}
W_{i \rightarrow j} e^{-\beta E_{i}} =W_{j \rightarrow i} e^{-\beta E_{j}}
\end{equation}

\begin{figure}[b]
\epsfbox{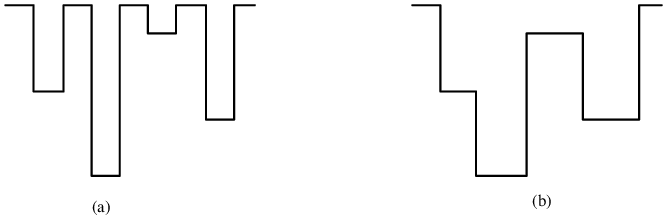}
\caption{(a):traps model; (b):``steps'' model; remember that the connectivity
is infinite, so that such a picture can be misleading!}
\label{fig}
\end{figure}

\noindent
For example, for the traps model, where the lifetime of configuration
$i$ is $\tau_{i} = exp(-\beta E_{i})$, the transition rates
are $W_{i \rightarrow j}=\frac{ exp(\beta E_{i})}{N}$.
Here we consider a transition probability depending on both $E_{i}$
and $E_{j}$, given by the Glauber dynamics:
\begin{equation}\label{glauber}
W_{i \rightarrow j}=\frac{1}{N} \ \frac{1}{1+exp(\beta (E_{j}-E_{i}))}
\end{equation}
As mentioned before, this system is quite different from the
trap model (see figure \ref{fig}). For instance, at
zero temperature, the jump to a lower state is allowed
in our model, while it is impossible to
jump out of a trap.

In this study, we will be interested in computing the law of diffusion
(the number of configurations reached at time $t$), the evolution of the
mean energy with time, the probability, given a
time $t$, to be in a configuration of lifetime $\tau$ (we shall define
the lifetime $\tau$ precisely below), which we will note
$p_{t}(\tau)$, and the two-times correlation function $C(t_{w}+t,t_{w})$.
This last quantity is defined as the mean overlap between the positions
of the system at times $t_{w}$ and $t_{w}+t$: the overlap is simply
either $1$ if the system is in the same configuration, or $0$ if it has
moved. Assuming an exponential decay out of the configurations, the
correlation function is related to $p_t(\tau)$ through:
\begin{equation}
C(t_{w}+t,t_{w})=\int_0^\infty d\tau \  p_{t_{w}}(\tau) e^{-t/\tau}
\end{equation}

Before turning to the exact solution of the master equation at zero
temperature, it is useful to start with a discussion of the trapping time
distributions.
When  the system is in configuration i, with energy $E_{i}$, the probability
of going away per unit time is
\begin{equation}
p_{s}(E_{i})=\sum_{j} \ W_{i \rightarrow j} \ .
\end{equation}
 For large N and using the definition
 (\ref{glauber}) of the transition probability, one gets at zero temperature:
\begin{equation}\label{unsurtau}
p_{s}(E_{i})= \int_{- \infty}^{E_{i}} dE' \ P(E')
\end{equation}
The ``trapping time'' $\tau_{i}$ is defined as
 $\frac{1}{p_{s}(E_{i})}$. It depends only on the energy of
the configuration, through the relation:
\begin{equation}
\frac{1}{\tau ^{2}}\frac{d\tau}{dE} = P(E) \ .
\end{equation}
We deduce that, {\it regardless of P(E)}, the a priori distribution of the
lifetimes is:
\begin{equation}\label{p0tau}
P_{0}(\tau) d\tau = \frac{d\tau}{\tau ^{2}}\theta (\tau - 1)
\end{equation}

As in the trap model, this is a broad distribution with a divergent mean
lifetime (although here it is  just marginally divergent, for any
$P(E)$). As was shown in \cite{Bouchaud1}, this fact in itself creates an aging
effect. In our case there is an additional effect because
the effective distribution of lifetimes evolves with time. After k jumps
the system will be in a lower energy configuration, and the probability
${\it P}_{k}(\tau)$ of having
a lifetime $\tau$ is different from $P_0(\tau)$. The zero temperature dynamics
gives the recursion relation:
\begin{equation}\label{rec}
P_{k+1}(\tau)=\it P_{0}(\tau) \ \int d\tau' \ \tau' P_{k}(\tau') \theta (\tau
-\tau ')
\end{equation}
which leads to:
\begin{equation}\label{pktau}
{\it P}_{k}(\tau)=\frac{(log \tau)^{k}}{k! \ \tau ^{2}}
\theta (\tau - 1) \ .
\end{equation}
The typical lifetime thus increases exponentially with $k$ (it means that
the diffusion is logarithmic). This will add up
to the usual effect of a diverging mean lifetime in order to induce aging.

We now proceed to the solution of the
master equation  at zero temperature using the Laplace transform.
Denoting by $p_{i}(t)$
the probability of being on configuration $i$ at time $t$, we have
\begin{equation}
\frac{d}{dt}p_{i}(t)=\sum_{j} \ T_{ij} \ p_{j}(t)
\end{equation}
with $T_{ij}=W_{j \rightarrow i}$ for $i \neq j$, and $\sum_{i} \ T_{ij}=0$.
The Laplace transform $\tilde p_{i}(\phi)= \int dt p_{i}(t) e^{-\phi t}$
satisfies the equation:
\begin{equation}\label{tl}
\tilde p_{i}(\phi)=\frac{p_{i}(0)+ \frac{1}{N} \sum_{j}\
 \theta ( E_{j} - E_{i} )
\tilde  p_{j}(\phi)}
{\phi + \frac{1}{\tau_{i}}}
\end{equation}
where the lifetime
$\tau_{i}$ is  defined as before by
$\frac{1}{\tau_{i}}= \sum_{j \neq i}\ T_{ji}$ and where we will
take $p_{i}(0)=\frac{1}{N}$.
To solve this equation we introduce the Laplace transform of the occupation
probabilities for all configurations of energy $E$:
\be
f(E,\phi) \equiv \sum_{j}  \tilde p_{j}(\phi) \delta (E-E_{j}) \ .
 \ee
Using (\ref{tl}) one derives for $f$ the equation:
\be
 \label{fg}
f(E,\phi)=g(E,\phi)\left( 1 + \int_{E}^{\infty} dE' f(E',\phi) \right)
\ee
where
\be
g(E,\phi)= \frac{1}{N} \sum_{j}\frac{\delta (E-E_{j})}{\phi
 + \frac{1}{\tau_{j}}}
=  \frac{ P(E)}{\phi + \frac{1}{\tau (E)}}
\ee
The self consistency equation (\ref{fg}) is easily solved and gives:
\be
f(E,\phi)=g(E,\phi) \exp \left( \int_{E}^{\infty} g(E',\phi) dE' \right) \ .
\end{equation}

We can now use this solution to compute the physical quantities of interest.
We start with the probability $p_t(\tau)$  to be a t time $t$ in
a configuration of lifetime $\tau$. Its Laplace transform with respect
to $t$ is given by:
\be
p_{\phi}(\tau)=\sum_{i} \tilde p_{i}(\phi) \delta (\tau - \tau_{i})
= \frac{P_{0}(\tau)}{\left( \phi + \frac{1}{\tau}\right)}
\frac{ f(E(\tau),\phi)}{g(E(\tau),\phi)}
 = \frac{\phi+1}{(1+\phi \tau)^{2}}
\ee
This Laplace transform can be inverted and gives:
\be
p_{t}(\tau)=\frac{t\tau -t + \tau}{\tau ^{3}} exp\left (-
\frac{t}{\tau}\right )\theta (\tau -1)
\end{equation}
We see that this expression decreases as $\frac{t}{\tau ^{2}}$ for
$t << \tau$, as for a model of traps for $P_0(\tau)=1/\tau^2$,
but the exponential term makes the probability
of being at time $t$ in a configuration with lifetime smaller than $t$ very
small.

The correlation function can also be computed for large $t$
and $t_{w}$:
\begin{eqnarray}
C(t_{w}+t,t_{w}) \simeq \frac{t_{w}}{t_{w}+t}
\end{eqnarray}
We see immediately the $\frac{t}{t_{w}}$ scaling of the
correlation function. The behaviour $ \lim_{t\rightarrow \infty}
C(t_{w}+t,t_{w}) =0$
is a consequence of the existence of dynamics even at zero temperature, while
the behaviour in
the other limit, $\lim_{t_{w}\rightarrow \infty} C(t_{w}+t,t_{w}) =1 $,
reflects the
existence of ``weak ergodicity breaking'' \cite{Bouchaud1} (the value $1$
for the limit is a consequence of $T=0$).

Let us emphasize that all  these results are independent of the distribution
$P(E)$. The two main hypotheses of the derivation are the fact that
the connectivity is infinite, and the temperature has been taken equal to zero.
These results can be partially extended in the case where $P(E)$ is an
exponential
distribution, $ P(E) \sim \rho \exp(\rho E) \theta (-E)$. Such a distribution
has been found
in mean field spin glass models \cite{mpv}, and it is at the heart of the
trap model description, since it leads to a broad distribution of lifetimes
when $\beta > \rho$.
Specializing to this case, we can first explain in more details the zero
temperature
dynamics studied above.
 Indeed, the relation
between energy and trapping time can be explicited as:
$
\tau_{i}=e^{-\rho E_{i}}
$
and the energy distribution evolves then as
\begin{equation}
{\it P}_{k}(E)=\frac{(-\rho)^{k+1} E^{k}}{k!} exp(\rho E) \theta (-E)
\end{equation}
The mean energy decreases as $-k/\rho$ with the number of
visited configurations, or equivalently as $-\log t/\rho$ (remember
that the diffusion is logarithmic).

For finite temperature, the set of self-consistent equations
is more complicated; we write $a_{i}=exp(\beta E_{i})$, so that
\begin{equation}
\frac{1}{\tau_{i}}= a_{i}^{\rho /\beta}\int_{0}^{a_{i}^{-\rho /\beta}}
\frac{dv}{1+v^{\beta / \rho}} \\
\end{equation}
and we obtain
\begin{eqnarray}
\tilde p_{i}(\phi)=\frac{p_{i}(0)+\int^{\infty}_{0} d\lambda
e^{-\lambda a_{i}} f(\lambda,\phi)}{\phi + \frac{1}{\tau_{i}}} \\
f(\lambda,\phi)=g(\lambda,\phi) + \int^{\infty}_{0} d\mu \
g(\lambda + \mu,\phi)f(\mu,\phi) \\
g(\mu,\phi)=\frac{1}{N}\sum_{j} \frac{a_{j}e^{-\mu a_{j}}}{\phi
 + \frac{1}{\tau_{j}}}= \frac{\rho}{\beta}\int_{0}^{1} du
\frac{ u^{\rho /\beta} e^{-\mu u}}
{\phi+u^{\rho /\beta}
\int_{0}^{u^{-\rho /\beta}}\frac{dv}{1+v^{\beta / \rho}}}
\end{eqnarray}

Taking $\lambda = \tau^{\beta /\rho}$ and writing $g(\tau ,\phi)$,
$f(\tau,\phi)$ instead of $g(\tau^{\beta/\rho},\phi)$,
$f(\tau^{\beta/\rho},\phi)$, we obtain for $f$ the following scaling:
\begin{equation}
f(\tau,\phi)= \frac{\rho}{\beta} \tau ^{1-\frac{\beta}{\rho}}
h(\phi \tau)
\end{equation}
with $h(x)$ behaving as $\frac{1}{x^{2}}$ for $x >> 1$ and
$\int_{0}^{\infty} dx h(x)$ finite. After some calculations, it can be
shown that $p_{t}(\tau)d\tau$ behaves like $\frac{d\tau \ t}
{\tau ^{2}}$ for
$1 << t << \tau$, and as $\frac{d\tau}{\tau}\left ( \frac{\tau}{t} \right )
^{1+ \frac{\beta}{\rho}} $ for $1 << \tau << t$, and for $\beta >> 1$.

We obtain thus qualitatively the same behaviour for $p_{t}(\tau)d\tau$
as before, and also for the correlation function: the dynamics is
not modified by a small temperature.

Note that this behaviour holds in the limit of infinite $N$; for any finite
$N$ the system eventually thermalises, after a time proportional to $N$
(for example, the minimal energy of $N$ states with exponentially distributed
energies
is $- \frac{log N}{\rho}$ so it takes a time $N$ to find it)

For this model,
some of our results are similar to those of \cite{Bouchaud3}: the mean energy
decreases as $-log(t)$, and at time $t$ the most probable configurations are of
lifetime $\tau = t$.
Nevertheless we must emphasize that the mecanism is totally different:
in a model of traps, the mean energy decreases because the system
visits more and more traps, and so it has more and more chances to
find deep ones. At each step ${\it P}_{k}(E)$ remains the same: $P_k=P_0$. The
diffusion is in $t^{\rho / 2\beta}$, and the energy at step k evolves
like $-log(k)$, because the minimum of $k$ energies distributed
according to the exponential distributionis in $-log(k)$.
In our model, on the opposite, the energy distribution that the particle
sees evolves, and so does the distribution of trapping times (see
(\ref{pktau}));
 the energy decreases in fact because it is
easier to find a configuration with lower energy (it takes a time
proportionnal to $exp(-\rho  E)$) than to move by thermal
activation (a time $exp(-\beta E)$ is needed). It means that the system
spends less time in a given configuration, but, since
the energy at step k decreases like
$-k$, the diffusion is much slower (logarithmic instead of a power-law),
so we finally get the same behaviour
for the mean energy as a function of time. However the main feature
is that there exists a zero-temperature dynamic, which is qualitatively not
modified by a small temperature.

It would of course be very interesting to be able to generalize
this approach beyond the case of an infinite connectivity. For instance
a more realistic definition of the REM dynamics could be to
start from the definition of the REM in terms of spins
with p ($\to \infty$) spin interaction \cite{derr,gm}, and use the
transition matrix resulting from single spin flip dynamics. This
seems rather complicated at the moment.

We thank J.P Bouchaud, R. Burioni, L. Cugliandolo, D. Dean
C. De Dominicis and J. Kurchan for many helpful discussions on related topics.

\end{document}